# Fulleryne, a new member of the carbon cages family

Mohammad Qasemnazhand[a], Farhad Khoeini[a*], Farah Marsusi[b]

[1]Department of Physics, University of Zanjan, P.O. Box 45195-313, Zanjan, Iran
[2]Department of Physics and Energy engineering, Amirkabir university of technology, P.O. Box 15875-4413, Tehran, Iran

**ABSTRACT**

In this study, based on density functional theory (DFT), we propose a new branch of pseudo-fullerenes which contain triple bonds with sp hybridization. We should call these new nanostructures fullerynes, according to IUPAC. We present four samples with the chemical formula of $C_{4n}H_n$, and structures derived from fullerenes. We compare the structural and electronic properties of these structures with those of two common fullerene and fullerene systems. The calculated electron affinities of the sampled fullerynes are negative, and much smaller than those of fullerenes, so they should be chemically more stable than fullerenes. Although fulleranes also exhibit higher chemical stability than fullerynes, but pentagon or hexagon of the fullerane structures cannot pass ions and molecules. Applications of fullerynes can be included in the storage of ions and gases at the nanoscale. On the other hand, they can also be used as cathode/anode electrodes in lithium-ion batteries.

**Introduction**

Carbon is an element that has the potential to adapt to different molecular structures and can form various molecular orbitals, such as sp, $sp^2$, $sp^3$, and so on. Diamond and graphite are the best-known carbon allotropes in bulk form that their structures are made of $sp^3$ and $sp^2$ hybridization, respectively. Recently, cumulene and carbyne have been introduced as new carbon allotropes having pure structures consisting of sp hybridization [1-7]. Some structures have more than one type of hybridization in their structures; for example, fullerene, which in addition to $sp^2$ hybridization, has a slight hybridization of $sp^3$ because of its curvature [8-11]. Graphyne is another example of new two-dimensional carbon materials that, unlike graphene, which includes $sp^2$ hybridization, also includes sp hybridization [12-17]. It has been suggested that graphene is transformed into fullerene [18], now the question comes to the mind that is there any material can be converted from graphyne?
In this study, we introduce four new structures of carbon cages that could be new branches of the pseudo-fullerenes family. The geometry of fullerene consists of twelve pentagon rings and a variable number of hexagons. If the structure of the fullerenes is saturated with hydrogen, the orbital hybridization of the bonds shifted from $sp^2$ to $sp^3$ that it is called fullerane [19].

In this work, we explore a new class of fullerenes derivatives by adding sp orbital hybridization to fulleranes structures. Then, by comparing the structural, electrochemical, and optoelectrical properties of these structures with fullerenes and fulleranes, we show that this a new class of carbon cages does not belong to the fullerenes and fulleranes categories. It is best to call them fullerynes in the style of IUPAC and the triple bonds with sp hybridization orbitals in their structures. We show that the fullerynes may have interesting applications in the field of nanotechnology [20], including hydrogen storage nanocapsules and cathode/anode electrodes in lithium-ion batteries.

**Computational Method**

We used four fullerenes and four fulleranes to start our study. Consider the structures shown in Fig. 1. The following structures belong to four fullerene classes, by saturating the following structures with hydrogen, their corresponding fullerane structures emerge.

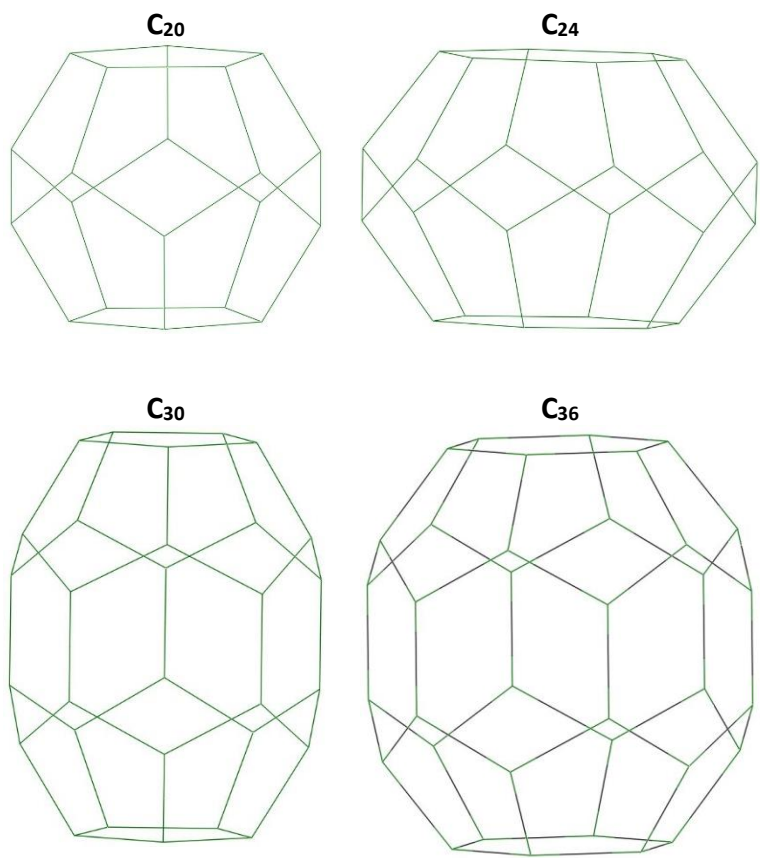

Fig. 1. The geometry of the fullerenes used.

We added two carbon atoms on each edge of the above structures, and we got four new structures. We must first determine that these four new structures are stable. We did this by calculating the infrared frequency of the optimized structures. Finally, we identify the structural, electrochemical and optoelectrical properties of the new cages and compare their properties with fullerenes and fulleranes.

The computation of the total energy, the optimum structures, and frequencies of the vibrations to check stability, has been performed using density functional theory (DFT). We used the B3LYP hybrid functional, which includes three parameters of Beck's correlation, and also includes Li, Yang, and Par electron exchange, though it consists of a portion of exchange from the Hartree-Fock (HF) method, too [21-22].

Elementary DFT method, underestimates the bandgap of the material, because it exaggeratedly predicts the density of occupied orbitals in wide location, on the other hand, the HF method, gives localized unoccupied orbitals, so it overrates the bandgap [23]. Therefore, the results of the hybrid functional are more consistent with the experimental results.

In our calculations, for describing the shapes of the orbitals, we used the lanl2dz basis set [24-27]. This basis set only imports valence electrons into the computation, so reduces computation time by freezing the inner electron shells. Our calculations are performed by the Gaussian 98 package [28].

**Results**

Fulleryne, the new carbon cage, is formed by adding two carbon atoms to each edge of the fullerane. So it can be concluded that in pure fullerynes twice the number of edges is added to the carbon number of each structure. In this study, the corresponding $C_{80}H_{20}$, $C_{96}H_{24}$, $C_{120}H_{30}$, and $C_{144}H_{36}$ fulleryne structures were obtained respectively for $C_{20}H_{20}$, $C_{24}H_{24}$, $C_{30}H_{30}$ and $C_{36}H_{36}$ fulleranes structures. In general, each fullerane with a chemical formula $C_nH_n$ has a fulleryne corresponding to the formula $C_{4n}H_n$. The carbon skeleton of the fulleryne structures obtained is shown in Fig. 2.

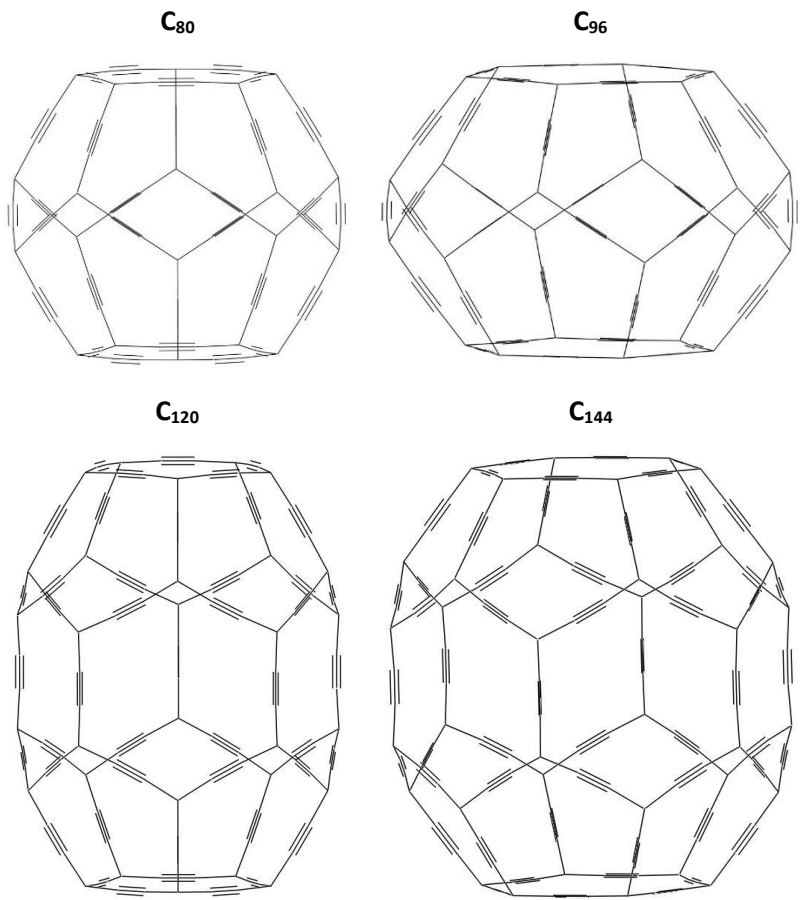

**Fig. 2. The obtained geometrical structure of the fullerynes.**

We obtained the infrared spectra of the presented structures to determine the stability of them and found that the frequency don't include imaginary in the vibrations [29]. The infrared spectra for the introduced fullerynes, presented in Fig. 3.

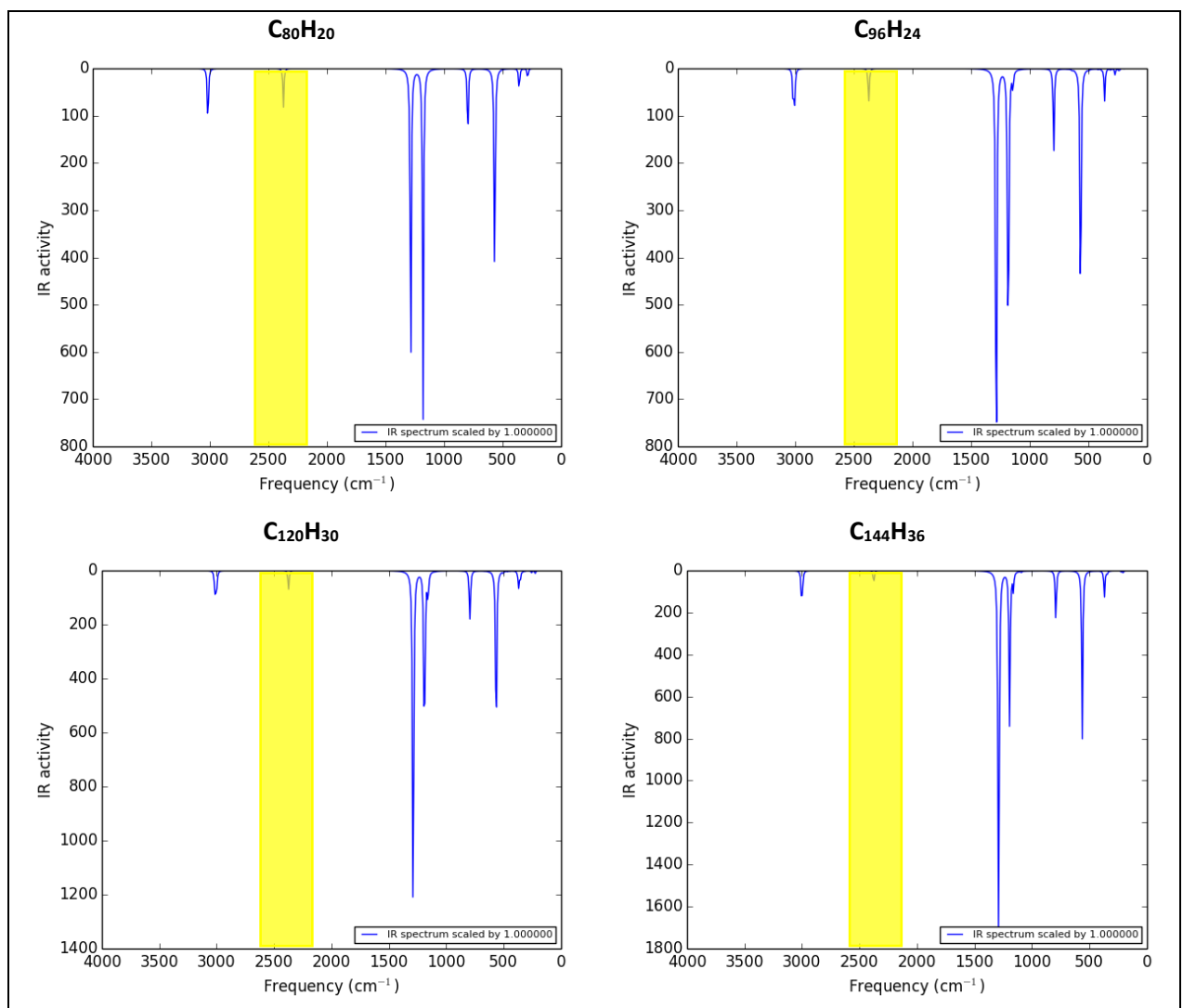

**Fig. 3.** The infrared spectra for the introduced fullerynes. The highlighted region is specific to the fullerynes.

By considering the above figure, we can understand the vibrational properties of the introduced fullerynes. In the diagram of the infrared spectrum, the absorption index in the range of 500 to 1000 is related to the carbon skeleton and in the range of 1000 to 1500 is related to the hydrogens of structure. The spectral region of these diagrams, which is specific to fullerynes, are frequencies of close to 2500, corresponding to the carbon added at the edge of the structures by sp hybridization, namely, the triple bonds.

After determining the stability of the fullerynes structures, we present fullerynes specifications in the following three parts: the structural, electrochemical and optoelectrical properties.

### i) Structural properties:

To describe the dimensions of the structures investigated in this study, it is assumed that each one is in a hypothetical fit box, and the dimensions of that box were obtained using a graphical interface, the GaussView software. Then the size of this box helps us to approximate the diameter of each structure. An example of a hypothetical box for the structure of $C_{20}$ is given in Fig. 4.

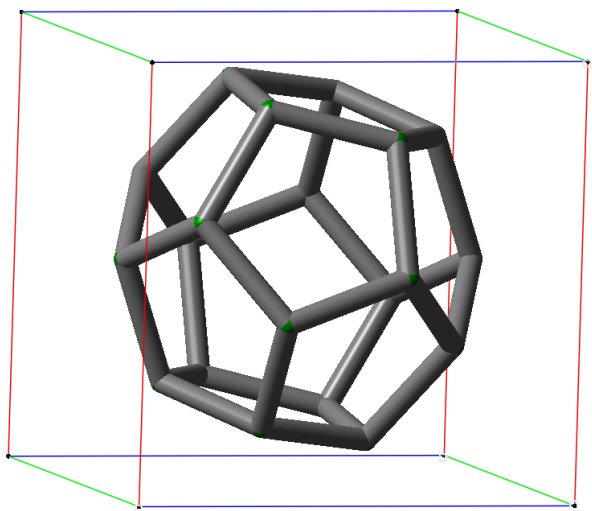

**Fig. 4. The $C_{20}$ structure inside a hypothetical box introduced by GaussView software.**

The size of the hypothetical box dimensions, along with the symmetry of each structure is given in Table 1. To imagine and compare the dimensions of the introduced fulleryne structures, the dimensional characteristics of the corresponding fullerane and fullerene structures are as follows:

Table. 1. Geometrical properties of the introduced fullerynes and the corresponding fulleranes and fullerenes.

| Cages | SYM | BOX (a, b, c) | | | $E_{tot}$ | BE |
|---|---|---|---|---|---|---|
| $C_{80}H_{20}$ | Ih | 14.6 | 14.6 | 14.4 | -83206.76 | 8.47 |
| $C_{96}H_{24}$ | D6d | 16.7 | 16.7 | 11.8 | -99848.07 | 5.96 |
| $C_{120}H_{30}$ | D5h | 18.0 | 15.1 | 14.8 | -124810.09 | 5.11 |
| $C_{144}H_{36}$ | D6h | 17.8 | 17.7 | 16.6 | -149772.02 | 5.11 |
| $C_{20}H_{20}$ | Ih | 7.5 | 7.5 | 6.7 | -21062.94 | 11.35 |
| $C_{24}H_{24}$ | D6d | 8.1 | 8.1 | 6.2 | -25274.61 | 11.30 |
| $C_{30}H_{30}$ | D5h | 8.4 | 7.7 | 7.5 | -31592.11 | 11.27 |
| $C_{36}H_{36}$ | D6h | 8.7 | 8.4 | 8.1 | -37908.18 | 11.20 |
| $C_{20}$ | Ih | 5.0 | 5.0 | 4.9 | -20714.29 | 7.49 |
| $C_{24}$ | D6d | 5.8 | 5.8 | 4.1 | -24859.11 | 7.57 |
| $C_{30}$ | D5h | 6.5 | 5.1 | 5.1 | -31080.30 | 7.79 |
| $C_{36}$ | D6h | 6.3 | 6.0 | 5.8 | -37305.00 | 8.03 |

The first column in Table 1 shows the type of structural symmetry. It can be seen from the above table that the symmetry of the corresponding structures is the same. However, the symmetry of the fullerane and fulleryne structures is higher than the symmetry of the fullerene structures, and by ignoring about 0.3 angstrom, the changes in the dimensions of the fullerenes can be classified into symmetrical groups of fulleranes and corresponding fullerynes. To illustrate this, it is necessary to examine the geometric properties of each structure, in detail. However, before that, for better visualization and comparison of these three types of carbon cages, the corresponding structures are shown in Fig. 5.

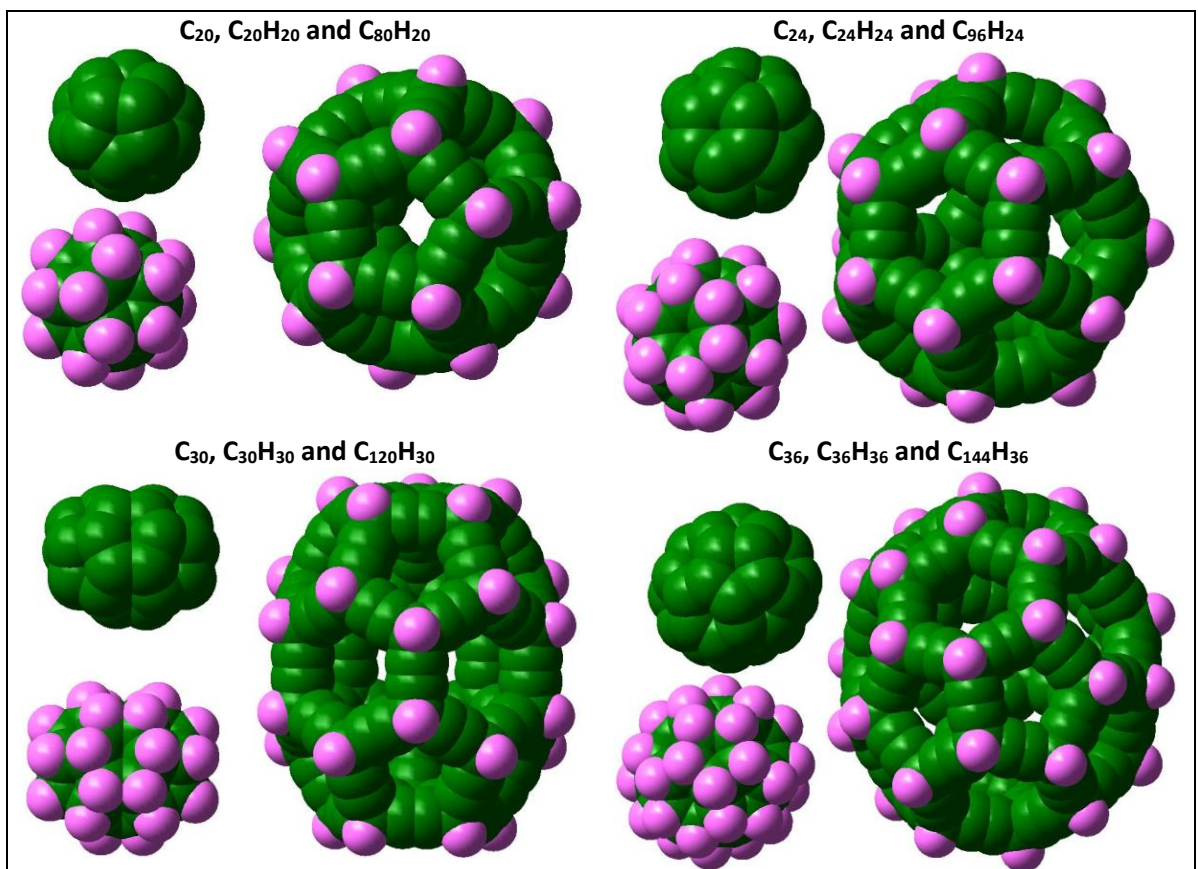

Fig. 5. Graphic models for fullerenes ($C_n$), fulleranes($C_nH_n$) and fullerynes($C_{4n}H_n$).

The introduced fullerynes in this study have two types of carbon bonds in their structures: The first type of bond is the bond between the vertex carbons and the edge carbons, and another is the carbon bonds located at the edges of the fulleryne structures, and this is a triple bond [30]. The lengths of the single and triple bonds of the investigated fullerynes are 1.49 and 1.22 angstrom, respectively. In fulleranes with the same symmetry group as fullerynes, the length of all carbon-carbon bonds is about 1.57 angstrom. While in fullerenes, the bond lengths are not uniform like those of fulleranes because of the resonance that makes the fourth electron [31]. For this reason, they are not symmetrical at the level of the fulleranes and the fullerynes, and with some slight exaggeration, they can fit into their symmetrical group. Using the colors, the bond length changes for each of the fullerenes, fulleranes, and fullerynes are schematically illustrated in Fig. 6.

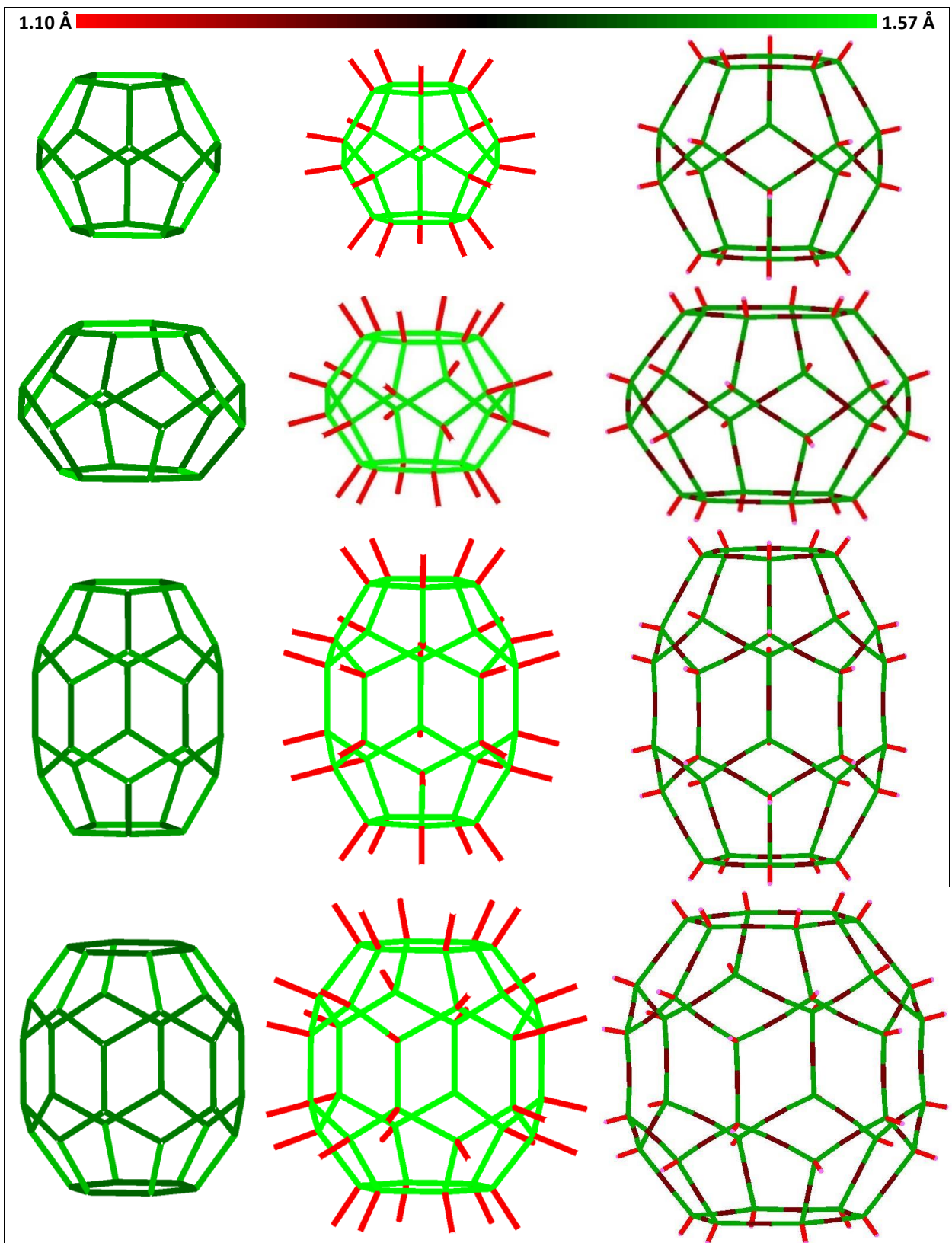

**Fig. 6.** Carbon-carbon bond length variations respectively in the three structures of fullerenes, fulleranes, and fullerynes.

The range of bond length changes starts from the lowest value of 1.10 Angstrom, which corresponds to the length of the carbon-hydrogen bond in fullerenes and fullerenes, and it continues to the maximum (1.57) related to carbon-carbon bonding in fulleranes. The red indicates the shortest bond length, the darker the middle bond length, and the light green the longer bond length.

Now we want to examine the stability of the cages. The total energy of each nanoparticle is related to the number of atoms forming it, so we can't use it as a factor for comparison, and use the binding energy, because it determines the contribution of a particle to the stability of the structure. We use the following equation to calculate the binding energy [32]:

$$BE = (E_{cage} - nE_{atom})/n. \tag{1}$$

Considering the binding energy with the nanocages, one can see that we are facing three different families of nanoparticles. Fulleranes with a binding energy of about 11 eV could be the most stable family of nanocages under study. On the other hand, the binding energy of the two groups of fullerenes and fullerynes is closer together and less than fulleranes. However, the fullerynes cannot be from the fullerenes family because the binding energy of the fullerynes, unlike the fullerenes, decreases as the nanocages grow larger, and in this case, it is similar to the fulleranes.

### ii)    Electrochemical properties:

Electrochemical properties are another factor that can be used to classify nanoparticles. The following table provides information on the energy levels of the HOMO and LUMO orbitals, the size of the HOMO-LUMO gap, and finally, chemical potential ($\mu$); in other words, same the Fermi energy level [33-34].

$$\mu = (E_{homo} + E_{lumo})/2. \tag{4}$$

We calculated the electronegativity ($\chi$), too, from the average amount in ionization potential and electron affinity. The following equations can be used to estimate the amounts of ionization energy and electron affinity [35]:

$$IP = -E_{homo}, \tag{2}$$

$$EA = -E_{lumo}. \tag{3}$$

We used the following relationships to obtain the chemical potential, the chemical hardness, and global softness:

$$\eta = (E_{HOMO} - E_{LUMO})/2, \tag{5}$$

$$\sigma = 1/\eta. \tag{6}$$

Now, with the help of the above equations, we calculate electrophilicity for different structures with the following relation [36]:

$$\omega = \mu^2 / 2\eta. \tag{7}$$

where it is a newer index and more accurately distinguishes between structures. The electron properties of the fullerynes introduced in this study are presented in Table 2. The following table also lists the electronic properties of corresponding fulleranes and fullerenes to illustrate the difference of fullerynes with them.

Table. 2. Electronic properties of the introduced fullerynes and the corresponding fulleranes and fullerenes.

| Fullerynes | HOMO | LUMO | H-L | μ | IP | EA | χ | η | σ | ω |
|---|---|---|---|---|---|---|---|---|---|---|
| $C_{80}H_{20}$ | -7.07 | 0.15 | 7.22 | -3.46 | 7.07 | -0.15 | 3.46 | 3.66 | 0.27 | 1.66 |
| $C_{96}H_{24}$ | -7.00 | 0.13 | 7.13 | -3.43 | 7.00 | -0.13 | 3.43 | 3.56 | 0.28 | 1.65 |
| $C_{120}H_{30}$ | -6.95 | 0.18 | 7.13 | -3.38 | 6.95 | -0.18 | 3.38 | 3.56 | 0.28 | 1.60 |
| $C_{144}H_{36}$ | -6.90 | 0.18 | 7.08 | -3.36 | 6.90 | -0.18 | 3.36 | 3.54 | 0.28 | 1.59 |
| **Fulleranes** | | | | | | | | | | |
| $C_{20}H_{20}$ | -7.08 | 2.01 | 9.09 | -2.54 | 7.08 | -2.01 | 2.54 | 4.54 | 0.22 | 0.71 |
| $C_{24}H_{24}$ | -6.99 | 1.73 | 8.72 | -2.63 | 6.99 | -1.73 | 2.63 | 4.36 | 0.23 | 0.79 |
| $C_{30}H_{30}$ | -6.60 | 1.43 | 8.03 | -2.58 | 6.60 | -1.43 | 2.58 | 4.02 | 0.25 | 0.83 |
| $C_{36}H_{36}$ | -6.48 | 1.09 | 7.57 | -2.69 | 6.48 | -1.09 | 2.69 | 3.78 | 0.26 | 0.96 |
| **Fullerenes** | | | | | | | | | | |
| $C_{20}$ | -5.90 | -4.14 | 1.76 | -5.02 | 5.90 | 4.14 | 5.02 | 0.88 | 1.14 | 14.32 |
| $C_{24}$ | -5.96 | -4.76 | 1.20 | -5.36 | 5.96 | 4.76 | 5.36 | 0.60 | 1.67 | 23.94 |
| $C_{30}$ | -5.92 | -4.66 | 1.26 | -5.29 | 5.92 | 4.66 | 5.29 | 0.63 | 1.59 | 22.21 |
| $C_{36}$ | -5.97 | -4.97 | 1.00 | -5.47 | 5.97 | 4.97 | 5.47 | 0.50 | 2.00 | 29.92 |

The above data clearly shows that we have three different spectra of nanostructures. The most chemical reactivity of the above structures are fullerenes, which have more electron electronegativity, chemical softness, and electrophilicity than fulleranes and fullerynes. As the size of the fullerenes increases, their reactivity increases too. Fulleranes, on the other hand, have the least reactivity, like the fullerenes, as their dimension's increase, their reactivity increases. Now by considering fullerynes properties, we see that their reactivity is between fullerenes and fulleranes, and unlike these two groups, as their dimensions grow, their reactivity decreases.

The density of the states diagrams also confirms the distinction between the three groups of fullerenes, fulleranes, and fullerynes. The density of states (DOS) diagrams for the investigated nanocages are obtained using GussSumm software [37]. DOS plots were presented in Fig. 7. Fullerenes have a larger electron affinity due to their negative lumo levels, but fulleranes do not tend to capture electrons due to positive lumo level. Although, the electron demand of the fullerynes is negative like that of the fulleranes, but not so strong because its negative LUMO level is close to zero.

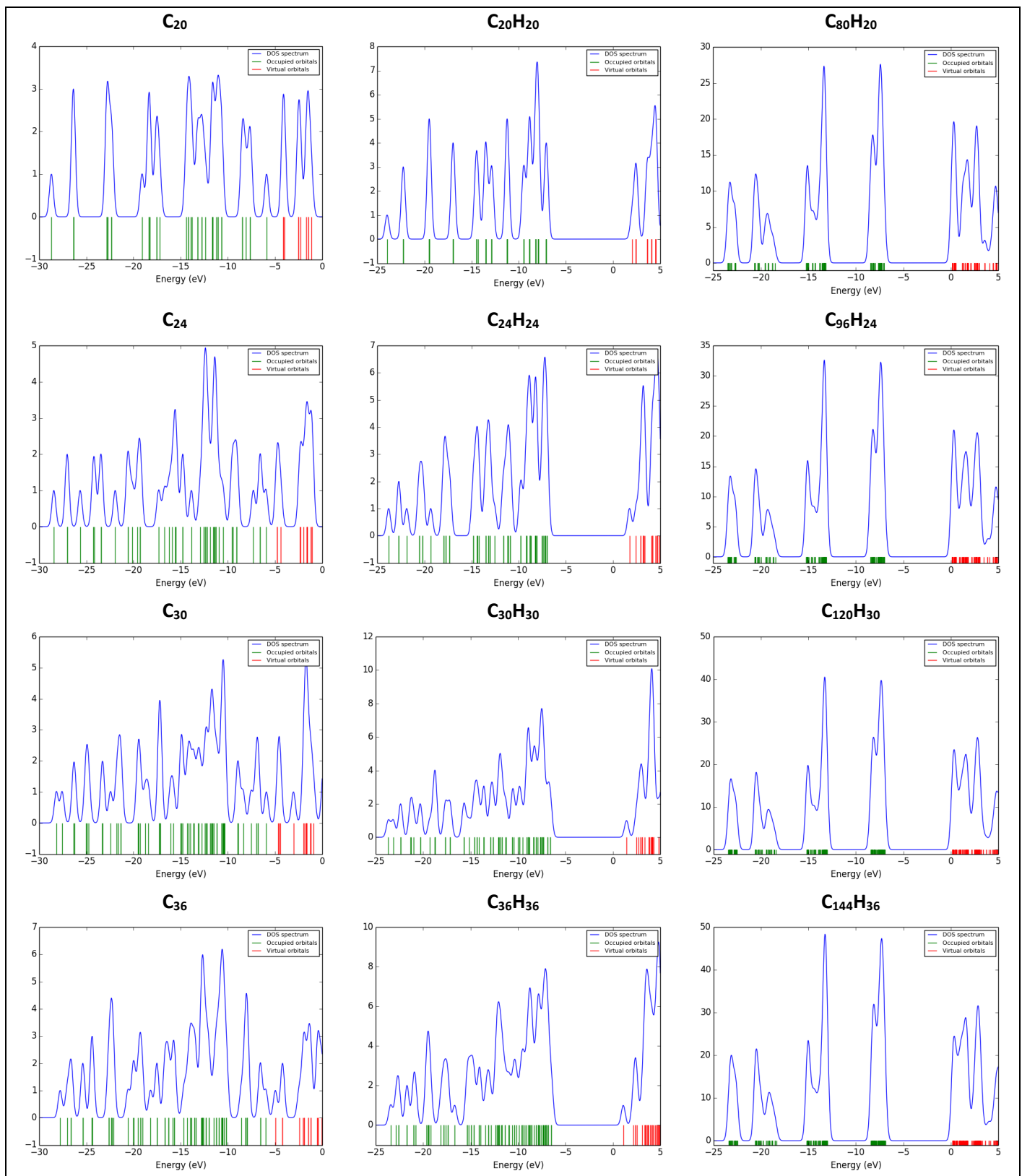

**Fig. 7. The density of the states' diagrams in the fullerene, fullerene, and fulleryne structures, respectively.**

### iii) Optoelectrical properties:

Other reasons why the fullerenes can be categorized into separate families can be mentioned differentiation of optoelectronic properties. In addition to the HOMO-LUMO gap, it is usually not equal to the absorption gap, and also, the absorption gap is not the same as the emission gap [38]. Schematically, the cause of the difference between the absorption and the emission gap is shown in Fig. 8.

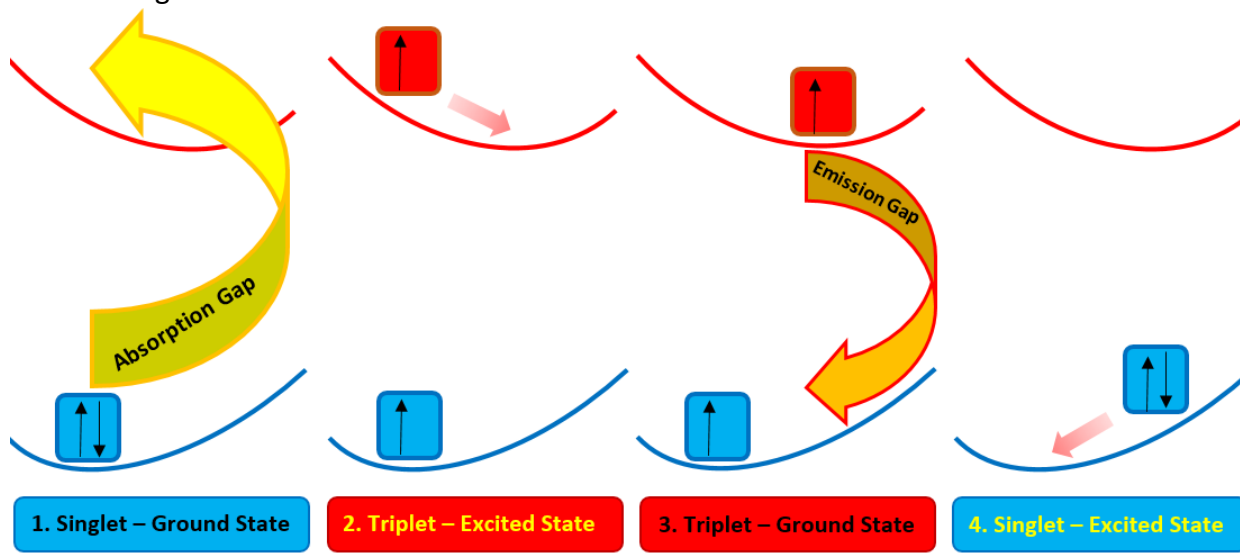

**Fig. 8. Schematic of the cause of the absorption and emission gap differences in the nanoparticles.**

We obtain the absorption gap for the investigated structures in this study by calculating the difference of total energy in the ground and excited states. To calculate the total energy for the excited state, we obtain the energy for the triplet state of the structures [39-42]. In Table 3, the total energy for the ground and the excited states, and finally, the absorption gap for the investigated structures are presented.

Table. 3. Absorption gaps of the introduced fullerynes and the corresponding fulleranes and fullerenes.

| Nanocages | Total energy | | Absorption gap |
|---|---|---|---|
| | Ground | Excited | |
| $C_{80}H_{20}$ | -83206.76 | -83201.24 | 5.52 |
| $C_{96}H_{24}$ | -99848.07 | -99841.74 | 6.33 |
| $C_{120}H_{30}$ | -124810.09 | -124804.51 | 5.58 |
| $C_{144}H_{36}$ | -149772.02 | -149765.33 | 6.69 |
| $C_{20}H_{20}$ | -21062.94 | -21054.80 | 8.14 |
| $C_{24}H_{24}$ | -25274.61 | -25266.81 | 7.8 |
| $C_{30}H_{30}$ | -31592.11 | -31584.92 | 7.19 |
| $C_{36}H_{36}$ | -37908.18 | -37901.34 | 6.84 |
| $C_{20}$ | -20714.29 | -20713.74 | 0 |
| $C_{24}$ | -24859.11 | -24859.12 | 0 |
| $C_{30}$ | -31080.30 | -31080.07 | 0 |
| $C_{36}$ | -37305.00 | -37305.22 | 0 |

Considering the above data, we conclude once again that we are facing three different families of nanocages. As can be seen, the quantum confinement effect (QCE) is observed in the size changes of the absorption band of the fulleranes [43], but in the fullerynes, there is no regular downward change in the absorption band. The fullerenes gap is so small that it is not common in optical works and is more suitable for electronic applications [44].

One of the most important reasons for the differences in the properties of these three nanocages is the difference in the situation of their electrons. To clarify this, we investigated the linear correlation ($Y=AX+B$) of the structural gaps with their dimensions using the model of a particle in the box [45].

$$\Delta E = \{const\}/m^* \cdot L^2 . \qquad (8)$$

where, we set the $\Delta E$ equivalent to the HOMO-LUMO gap and obtain $1/L^2$ from the values in Table 1. Finally, we obtained for fullerene, $Y=157.57X-0.47$, for fullerane, $Y=586.99X+4.96$, and for fulleryne, $Y=266.99X+6.78$, as linear relation. We conclude that the effective mass of the electron is different in the three groups of fullerenes, fulleranes, and fullerynes. Since the slope of the line of relation to the fullerene is lower than the others, the heaviest effective mass is related to its electron. So, according to Fig. 8, we expect more stokes shift for the fullerane.

## Application

Fulleryne's chemical stability makes it suitable for storing and transporting some ions or gases at the nanoscale. Although fulleranes also exhibit higher chemical stability than fullerynes, but pentagon or hexagon of the fullerane structures cannot pass ions and molecules. In this section of the report, we have examined the results of the interaction of the fullerene structure with lithium ion. Using the following relationship, we can find the absorption energy ($E_{abs}$) between ions and structures such as cages [46-47].

$$E_{abs} = E_{ion} + E_{cage} - E_{total} . \qquad (9)$$

The following table (Table.4) shows the results of the calculations of the absorption energy between the fulleryne cage ($C_{80}H_{20}$) with the atom and the lithium-ion in the center positions of the cage and its face.

Table. 4. Table of Energy, X@$C_{80}H_{20}$ means the particle is in the center of the cage, and X_$C_{80}H_{20}$ means the particle is in the face of the cage.

| Cage+X | $E_X$ | $E_{Cage}$ | $E_{total}$ | $E_{abs}$ |
|---|---|---|---|---|
| Li@$C_{80}H_{20}$ | -203.84 | -83209.81 | -83413.63 | -0.02 |
| Li_$C_{80}H_{20}$ | -203.84 | -83209.81 | -83413.34 | -0.31 |
| Li$^+$@$C_{80}H_{20}$ | -198.24 | -83209.81 | -83409.52 | 1.47 |
| Li$^+$_$C_{80}H_{20}$ | -198.24 | -83209.81 | -83410.55 | 2.50 |
| H$_2$@$C_{80}H_{20}$ | -31.43 | -83209.81 | -83241.24 | -0.00 |
| H$_2$_$C_{80}H_{20}$ | -31.43 | -83209.81 | -83241.07 | -0.17 |

MD calculations show that fullerene cages can serve as a nanoscale reservoir for hydrogen gas [20], to find out how the ions are released into the fullerene cage; we compared their absorption energy with the hydrogen absorption energy in the cage, so that we can see the values in the table. Our calculations show that the lithium atom, like a hydrogen molecule, can move freely in the fulleryne cage, while the lithium ion is bound to the fulleryne cage, especially in its face. This feature makes the fulleryne cage susceptible to use as a cathode/anode electrode in lithium-ion batteries. A schematic of the process of lithium-ion reduction and releasing lithium atom from the fulleryne cage is shown in Fig. 9.

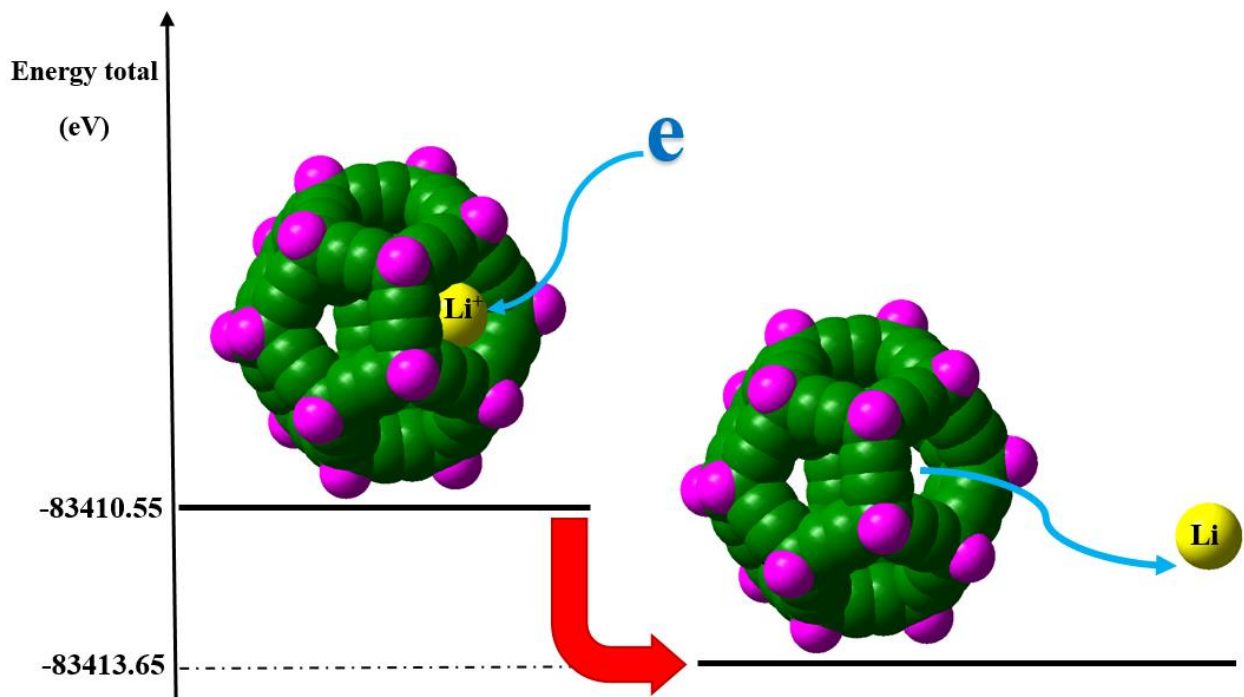

Fig. 9. Schematic of the lithium atom inside the $C_{80}H_{20}$ fulleryne cage.

According to the above figure and Table 4, it can be noticed that by oxidation of lithium atom, the energy level of the system increases (Discharge the battery).

## Summary

In this work, we have introduced a new branch of carbon nanocages; it is best to call them fulleryne, in the style of IUPAC, because of the triple bonds that exist in its structure. Fulleryne's chemical stability makes it suitable for storing and transporting some ions or gases at the nanoscale. Examination of the electrical, structural, and optical properties shows that the fullerynes fall into a category independent of known carbon cages.

**Conflicts of interest**

There are no conflicts to declare.

*Corresponding author's email: khoeini@znu.ac.ir